\documentclass[prb,twocolumn,showpacs,superscriptaddress]{revtex4}
\usepackage{graphicx}
\usepackage{dcolumn}

\begin{document}

\title{Steering effects on growth instability during step-flow growth
of Cu on Cu(1,1,17) }

\author{Jikeun Seo}
\affiliation{Division of General Education, Chodang University,
 Muan 534-701, Republic of Korea }
\author{Hye-Young Kim}
\affiliation{Department of Chemical Engineering, The Pennsylvania
 State University, University Park, PA 16802}
\author{J.-S. Kim}
\affiliation{Department of Physics, Sook-Myung Women's University,
 Seoul 140-742, Republic of Korea}

\date{\today}

\begin{abstract}

 Kinetic Monte Carlo simulation in conjunction with molecular dynamics
 simulation is utilized to study the effect of the steered deposition
 on  the growth of Cu on  Cu(1,1,17). It is found that the deposition
 flux  becomes inhomogeneous in step train direction and the
 inhomogeneity depends on the deposition angle, when the deposition is
 made along that direction. Steering effect is found to always increase
 the growth instability, with respect to the case of homogeneous
 deposition. Further, the growth instability  depends on the deposition
 angle and  direction, showing minimum at a certain deposition angle
 off-normal to (001) terrace, and shows a strong correlation with the
 inhomogeneous deposition flux. The increase of the growth instability
 is ascribed to the strengthened step Erlich Schwoebel barrier effects
 that is caused by the enhanced deposition flux near descending step
 edge due to the steering effect.

 \end{abstract}


 \maketitle

 \section{Introduction}

 In the growth of thin films on a vicinal surface of high areal step density,
 there is a net current of deposit particles towards the ascending step edge
 due to the Erlich Schwoebel barrier at descending step edge.
 Such transport of deosit atoms to the ascending step increases the
 possibility  of step flow growth and makes the growth of thin films
 on a vicinal surface more stable than that on a sigular
 surface.\cite{vici0,Politi}  Moreover, such asymmetric flow of deposit
 atoms  provides the possibility of forming a structure  along the step
 edge and has been a subject of numerous studies for the growth of
 one-dimensional systems.\cite{line-by-line}  Even in the thin film
 growth on a vicinal surface, however, develops the meandering
 instability  along  step-edge.\cite{Bail, Ernst1, Ernst2}
 Possible sources for this instability\cite{Politi} have been suggested
 as the asymmetric adatom diffusion due to the step Erlich-Schwoebel
 barrier \cite{Bail, Ernst2} and kinetically limited diffusion of
 the adatoms due to the  kink Ehrlich-Schwoebel
 barrier.\cite{vici1, Koponen} \par

 In addition to the kinetic effects mentioned above, the deposition
 process, one of the ignored dynamic processes,  has been recently
 found to affect the thin film growth\cite{steering1}.  That is, the
 interaction between a deposit atom and the atomic structure on the
 surface modifies   the trajectory of the deposit atom, called
 steering effect, and causes the inhomogeneous distribution of adatoms
 affecting the growth of thin films.\cite{steering1, steering2}
 Adjacent to the edge of  islands or steps, the steering effect is
 conspicuous due to rapid variation of the interaction potential, and thus
 is expected to be more influential in a deposition on a vicinal surface
 having high areal step density than in that on a singular surface.\par

 The purpose of the present study is to explore the role of steered
 deposition on the thin film growth on a vicinal surface, which has
 been ignored in most of the previous simulation or theoretical
 studies (see Ref. 2 for a review).  Specifically, we calculate the
 deposition flux distribution on the vicinal surface,
 varying the deposition angle and direction. We also search for any
 possibility to overcome such kinetic growth instability by adjusting
 the dynamic variables involved in the deposition process. We have
 chosen to study the growth of Cu on  Cu(1,1,17), because Cu(1,1,17)
 shows no surface reconstruction and has been a subject of
 many experimental and theoretical studies\cite{Koponen,Nissila}
 allowing us to compare our results with preexisting ones.
 Present study utilizes a computer simulation combining a molecular
 dynamics (MD) simulation for the dynamics of deposit atoms with a
 kinetic Monte Carlo (KMC) simulation for the growth of adatoms on
 the surface.\par

 We find that the steering-induced enhancement in the deposition flux
 near descending step edge is a critical factor affecting the growth
 instability on vicinal surface. The inhomogeneity of deposition flux
 depends on deposition angle, and a deposition angle which gives the
 minimum growth instability is found. Nevertheless, the steering effect
 always  increases the growth instabililty regardless
 of the deposition angle, with respect to the case where steering effect
 is neglected. \par

 \section{Simulation method}

 KMC simulation is adopted to simulate the whole process of thin film
 growth. Contrary to conventional KMC scheme,to simulate the trajectories
 of depositing atoms in detail, we incorporate MD
 into KMC simulations, where MD is employed  whenever a deposition
 event occurs in the KMC\cite{steering2}. \par

%%%%%%Fig. 1-diffusion processes%%%%%%%%%%%%
 \begin{figure}[pb]
 \includegraphics[width=0.45\textwidth]{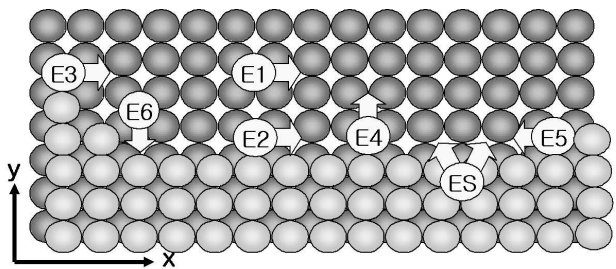}
 \caption{Illustration of some diffusion processes
taken into account in the present simulation.}
 \end{figure}
 %%%%%%%%%%%%%%%%%%%%%%%%%%%%%%%%

%%%%%%%Table I-Diffusion barriers and diffusion parameters %%
 \begin{table}
  \caption{\label{table1} Diffusion barriers and parameters used in KMC.
 Same notation is used for each diffusion process as in Fig. 1. }
  \begin{ruledtabular}
  \begin{tabular}{ccc}
  diffusion type & diffusion barrier \\
  \tableline
  E1 & 0.42 eV  \\
  E2 & 0.38 eV  \\
  E3 & 0.51 eV \\
  E4 & 0.68 eV   \\
  E5 & 0.59 eV   \\
  E6 & 0.18 eV   \\
  ES & E1+0.1 eV  \\
  \tableline
  jump frequency($\nu_{0}$) &  $3.6 \times 10^{12} $ \\
  deposition rate ($F_{0}$) &  0.003 ML/s \\
  \end{tabular}
  \end{ruledtabular}
  \end{table}
  %%%%%%%%%%%%%%%%%%%%%%%%%%%%%%%%%%%

 In the MD simulation, a Lennard-Jones potential
 $U(r) = 4D[ (\sigma /r)^{12}- (\sigma /r)^{6}]$
 is used for the pair interaction between a deposit atom and an atom
 on surface,  with $D=0.4093$ eV and $\sigma= 2.338 \AA$. These values
 of $D$ and $\sigma$ are adopted from Dijken $et. al.$ \cite{steering1,
 Sanders}. The initial kinetic energy of the deposit atom is set to
 0.15 eV, corresponding to the melting temperature of Cu. The
 Newton's equation of motion is solved using Verlet algorithm.
 Atom is approached to  the substrate by MD, and then positioned
 to the nearest four-fold hollow site from the terminal position.
 The transient mobility  is not included in the present study.
 That is, the deposit atoms are assumed to be in equilibrium
 with the substrate right after the deposition.\par

 In the KMC simulation, a lattice gas model is adopted which allows
 jump diffusion for adatom motion on fcc lattice. The possibility of
 each jump diffusion is calculated  from the corresponding hopping rate,
 $\nu = \nu_{0}  \exp^{-\beta E}$,  with attempt frequency,
 $\nu_{0}= 3.6 \times 10^{12} /s$. The definitions of the most relevant
 diffusion processes in the present simulation are illustrated in Fig. 1.
 In Table I, listed are the values of the diffusion barriers, $E_i$,
 that are adopted from the values used by Koponen\cite{Koponen, Nissila}
 in a growth simulation of Cu on Cu(1,1,17) and those obtained by
 Furmann\cite{Zuo} from a simulation study for thin film growth on
 Cu(001). \par

 Cu(1,1,17) surface has a (001) terrace of 8.5-atomic width between
 two steps of an atomic height in [-1,1,0] direction. In the following,
 x-axis is along the step edge as shown in Fig. 1, and y-axis is along
 the step train direction.  The simulation  box has 12 terraces with
 step edge length of 800 $a_{0}$, where  $a_{0}$ is the surface lattice
 constant of Cu(001), 2.55 $\AA$. Periodic boundary conditions are
 adopted in both x and y directions. \par

 \section{Results and Discussion}

 %%%%%%Fig. 2%%%%%%%%%%%%%%%%%%%%%%
 \begin{figure}
 \includegraphics[width=0.45\textwidth]{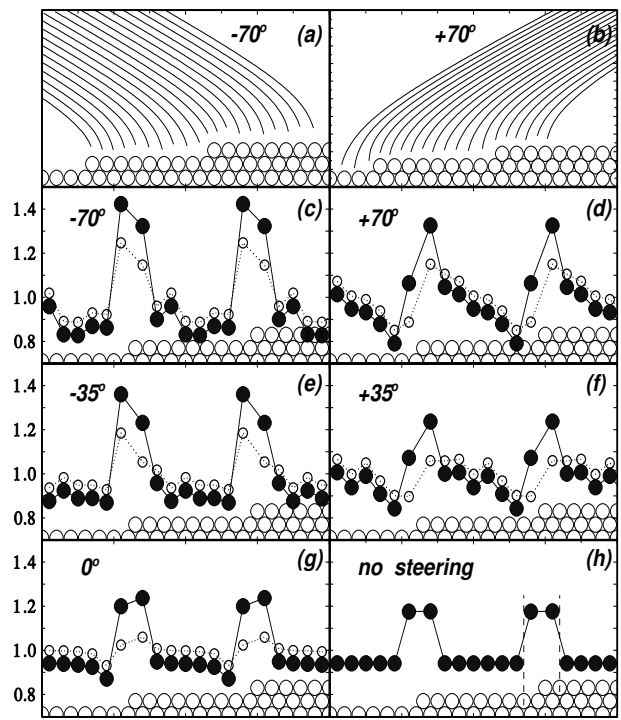}
 \caption{Trajectory of deposit atoms and normalized deposition flux.
 (a) Trajectory of deposit atoms at deposition angles of (a) $-70^o$
 and  (b)$+70^o$. Steered deposition fluxes at deposition angles
 of (c) $-70^o$ , (d)$+70^o$,  (e) $-35^o$ , (f)$+35^o$, (g) $0^o$,
 and (h)  deposition without steering effect.
 Normalization is made with respect to homogeneous flux.  Solid circle:
 Deposition flux. Open circle : Deposition flux after subtracting the
 enhancement due to the purely geometrical contribution (see the main
 text for details).}
 \end{figure}
 %%%%%%%%%%%%%%%%%%%%%%%%%%%%%%%%

 As a preliminary investigation of the steering effect on thin film
 growth,  the deposition flux distributions or deposition probabilities
 are examined for  various deposition angles by MD.  Deposition
 angle is measured from the normal to the (0,0,1) terrace to [-110]
 direction (y-axis).  The positive deposition angle is for the deposition
 direction from the upper terrace to the lower one along y-axis as shown
 in Fig. 2(b),
 and the negative angle is for the opposite direction as shown in Fig. 2(a).
 The trajectories of the deposit atoms in Figs. 2(a) and (b) show the
 steering effect,  where bending of trajectories of the incident atoms,
 most notably near steps, occurs due to the interaction between the
 deposit atom and substrate atoms.  Figs. 2(c) to (g) show the deposition
 flux distributions normalized to homogeneous flux for various deposition
 angles. Depositon flux, shown with solid circles in Figs. 2(c) to (g),
 increases near step , while that on terrace decreases
 compared with homogeneous flux.\cite{size-dependence}
 It is important, however, to note that this enhanced flux near step is not
 soley due to the steering effect. The deposition flux distribution in
 Fig. 2(h) is for deposition with no steering effect considered,
 and still shows relatively high flux near steps. This is because there
 are only two adsorption sites available in 2.5 $a_{0}$ distance from
 each step edge along y-axis, while one adsorption site is available
 in each 1.0 $a_{0}$ distance on terrace.  The deposition flux after
 subtracting this purely geometrical contribution is
 shown with open circles in Figs. 2(c)  to (g) and shows the enhanced
 deposition flux near steps purely due to the steering effect. \par

 For deposition angles closer to the grazing angle (that is,
 anlges of larger magnitude), the deposition flux becomes more
 inhomogeneous or more enhanced near steps, as can be seen by
 comparing Figs. 2(c) and (d) with Figs. 2(e) and (f), respectively.
 As deposition angle becomes larger, so does the flight time of
 depositing atoms, during which their trajectories and in turn,
 the deposition fluxes are apt to be more disturbed by the
 inhomogeneous substrate potential. It is also interesting to note
 the difference between the flux profiles at positive deposition
 angles (Figs. 2(d) and (f)) and those at negative deposition
 angles (Figs. 2(c) and (e)). In the negative angles, the deposition
 flux at the ascending step edge is larger than
 that at the descending step edge, and {\it vice versa}.  This may be
 explained from the fact that at positive deposition angles, the
 shadowing effect\cite{steering2} diminishes the probability for
 deposit atoms to sit on the sites next to the ascending step edge,
 while no such  shadowing is expected for negative
 deposition angles.\par

%%%%%%Fig. 3%%%%%%%%%%%%%%%%%%%%%%
 \begin{figure}
 \includegraphics[width=0.45\textwidth]{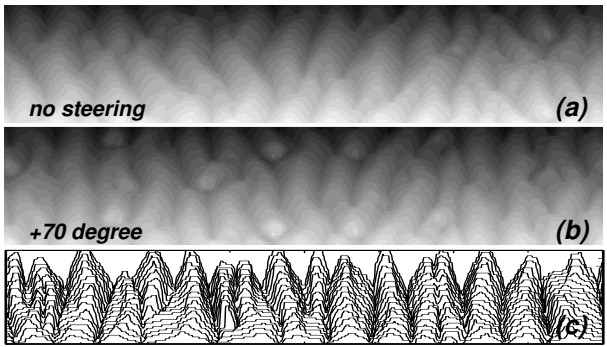}
 \caption{
Snapshot images of the 5ML Cu grown on Cu(1,1,17) at 240K.
(a) Deposition with no steering effect.
(b) Steered deposition at   $70^{o}$.
The size of the figures, (a) and (b), is 800$\times$150 $a_{0}^{2}$.
 (c) Evolution of a step edge with
increasing coverage. Successive curves show the development of step edges
at Cu coverages below 5 ML with the increment of 1/3 ML. The size of
the figure (c) is 800$\times$54 $a_{0}^{2}$.
$a_{0}$ is the surface lattice constant of Cu(001), 2.55 $\AA$.}
 \end{figure}
 %%%%%%%%%%%%%%%%%%%%%%%%%%%%%%%%

 The effect of the steering-induced inhomogeneous deposition flux
 on thin film growth on a vicinal surface is studied by KMC utilizing
 MD for each deposition event. During the growth, the substrate
 temperature is set to 240 K. Snapshots of a simulated system are
 shown in Figs. 3(a) and (b).  Fig. 3(c) shows the  evolution of
 a step as the coverage increases.  We observe that the average
 position  of step edge proceeds 8.5 $a_{0}$ for each monolayer (ML)
 deposition, indicating  step flow growth.  However, the lateral
 roughness increases, and the coherence between adjacent step
 edges develops  to form 'finger'-like structures  (Fig. 3(c))
 as the coverage increases. Each finger shows  ledge envelope
 along [100] and [0,-1,0] directions, as observed for both
 experimental studies\cite{Ernst2} and the simulation results by
 Koponen {\it et al.}.\cite{Koponen} \par

 For a quantitative understanding of the growth instability on a
 vicinal surface, the lateral roughness and the finger width taken
 as a measure of lateral coarseness are calculated.  We define the
 lateral height, $h(x)$, as the distance from a position x at a
 pristine step edge  to the growth front in the direction normal
 to the step edge (that is, in y-direction), and the lateral
 roughness as $w(x) \equiv \sqrt{<{{h(x)}^{2}-<h(x)>^{2}}>}$.
 The lateral coarseness is calculated from the average separation
 between fingers within heights $h_{avg} \pm 5 a_{0}$, where $h_{avg}$
 is the average lateral height of each step. As a measure for the growth
 instability, we take the aspect ratio, lateral roughness to finger
 width (lateral coarseness).  For an ideal step-flow growth or a stable
 growth, the aspect ratio should be very small.\par

%%%%%%Fig. 4%%%%%%%%%%%%%%%%%%%%%%
 \begin{figure}
 \includegraphics[width=0.45\textwidth]{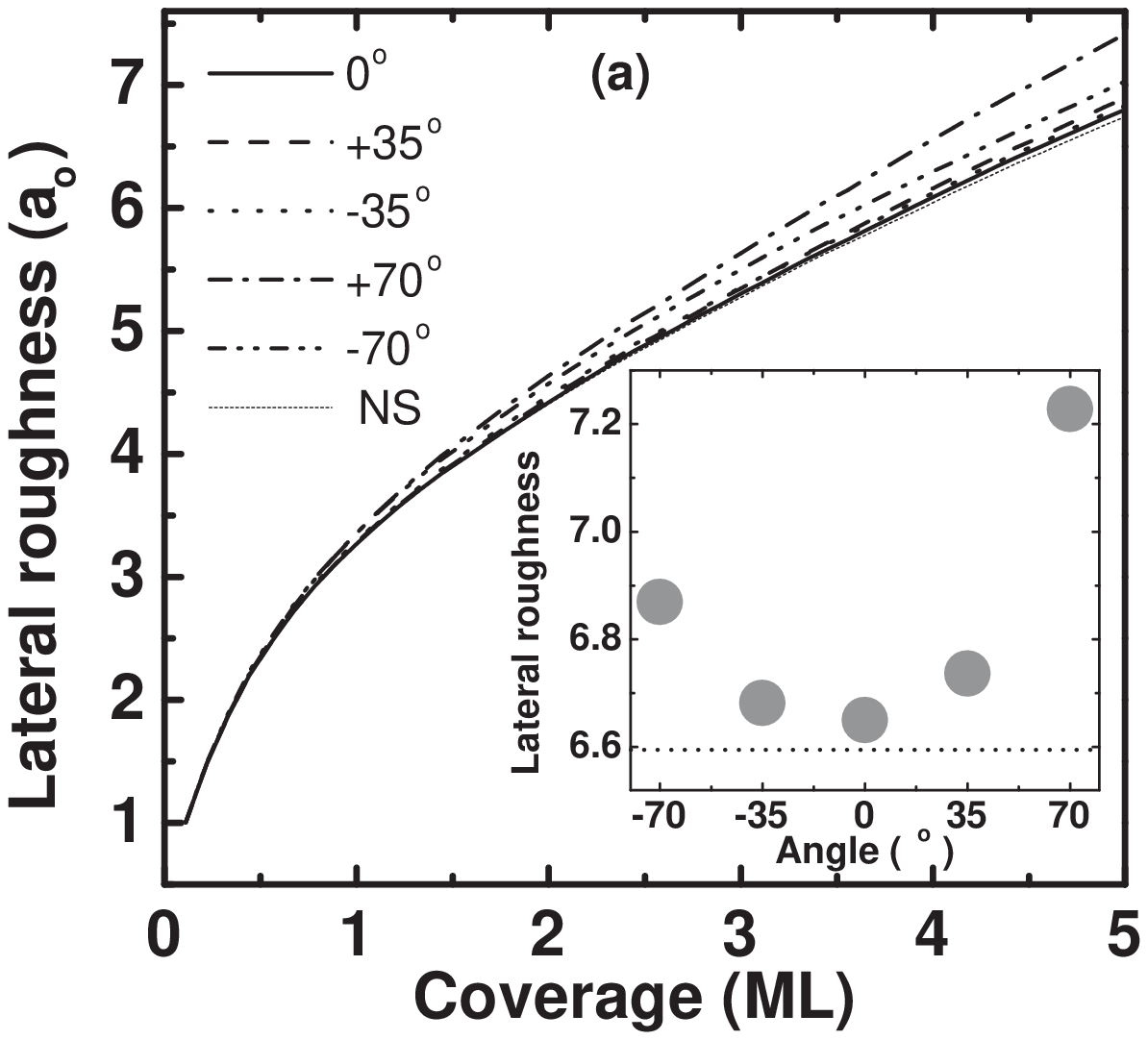}
 \includegraphics[width=0.45\textwidth]{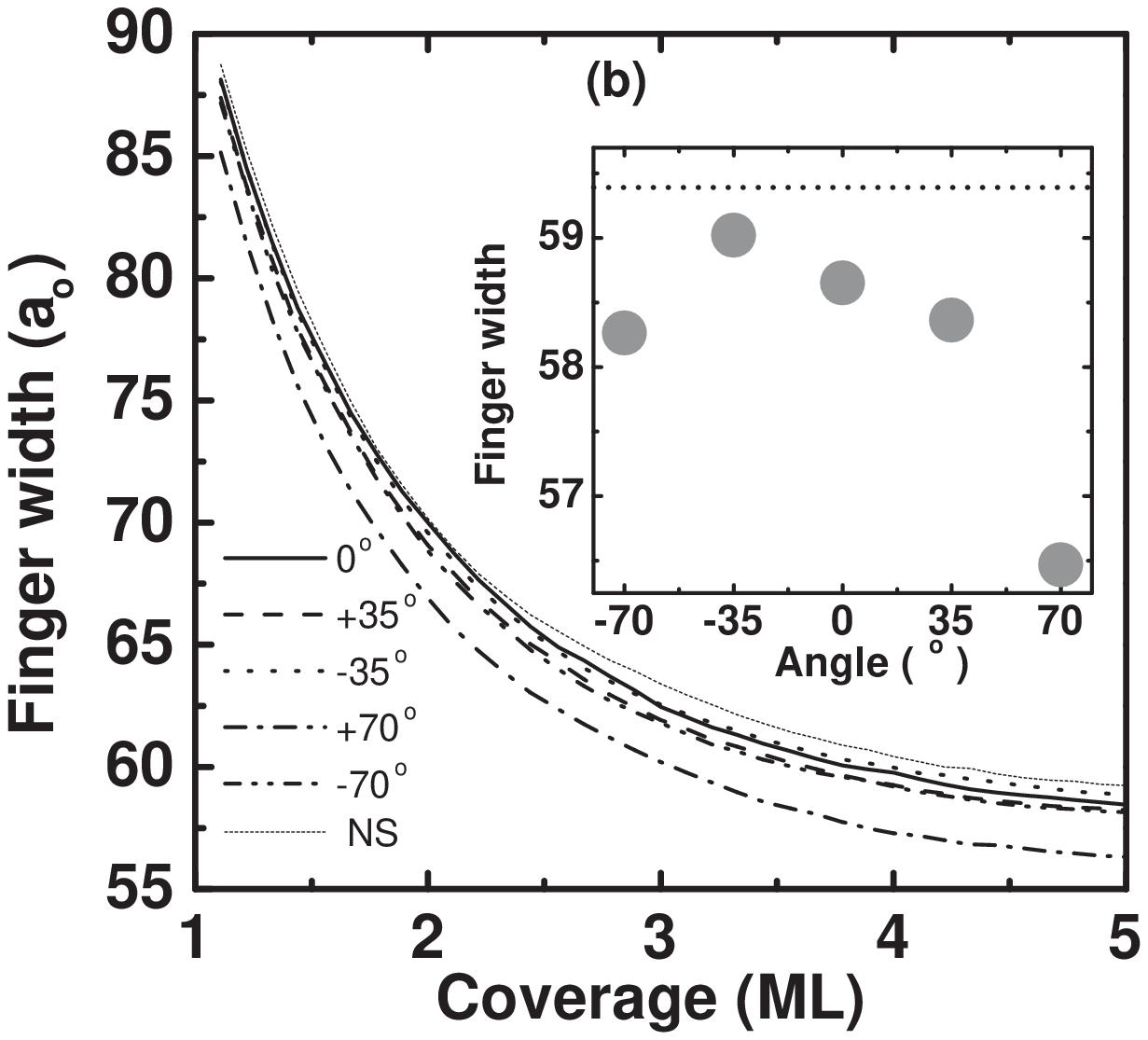}
 \caption{ (a) Lateral roughness and (b) finger width (lateral coarseness)
 as function of coverage in the growth of Cu on Cu(1,1,17) at 240K. Refer
 main text for the definitions of lateral roughness and finger width.
 Inset: (a)  lateral roughness and (b) finger width  as a function of
 deposition angle after depositing 5 ML.
The dotted lines in the figures (NS) and insets
 are the results of growth without considering  the steering effect. }
 \end{figure}
 %%%%%%%%%%%%%%%%%%%%%%%%%%%%%%%%

 In Fig. 4(a), the lateral roughness increases monotonically as
 coverage increases. At the maximum coverage of the present
 simualtion, 5 ML, the roughness is about 7$a_{0}$ indicating
 a very rough step edge. The roughness  shows distinct dependence
 on the deposition angle. In  the inset of Fig. 4(a),  shown is the
 lateral roughness as a function of deposition angle after depositing 5 ML.
 The roughness is minimum at deposition angles at ${\rm 0^o}$.
 As the deposition angle becomes larger, so does the roughness.
 In addition to the deposition angle, the roughness depends also
 on the direction of deposition. When deposition is made facing
 ascending step edge or at negative angle, the roughness of the
 film is small compared with that grown at the same magnitude of
 depostion angle, but in the opposite direction facing descending
 step edge.\par

  The development of lateral coarseness with increasing
 coverage was estimated by that of finger width. In Fig. 4(b)
 and its inset, the finger width monotonically dereases as coverage
 increases, and also shows a definite dependence on both deposition
 angle and direction. The finger width shows maximum at
 ${\rm -35^o}$ and decreases to minimum at ${\rm + 70^o}$.
 The most notable thing is that the lateral roughness and coarseness
 have close correlation in their  dependence on the deposition angle ;
 deposition at angles between  ${\rm - 35^o}$ and ${\rm 0^o}$
 shows most stable step-flow growth with the  minimum roughness
 and maximum finger width or the minimum aspect ratio, while
 deposition at $+ 70^o$ shows the opposite behavior, the most
 unstable growth with the maximum roughness and the minimum finger width,
 or the maximum aspect ratio.\par

 The aforementioned angular dependence of the growth instability should
 have originated from the dynamic effect of deposition process, the
 steering effect, since all the kinetic variables are identical
 for each deposition at various angles.
 A direct result of steering effect is the inhomogeneous deposition flux.
 Hence, we investigate the correlation between depsoition flux distribution
 and growth stability: The atoms deposited  near ascending step edge is
 expected to reproduce the step edge by directly adhering to the sites
 near step edge, and should not be the main source of steering-induced
 growth instability.  However, the atoms near the descending step edge
 would diffuse across terrace before reaching ascending step edge due
 to step Erlich Schwoebel barrier. During such terrace diffusion,
 the atoms redistribute themselves to feed and newly form laterally
 inhomogeneous structures, being a source of meandering
 instability.\cite{Bail}  Indeed, we find the predicted correlation
 between the growth instability and the enhanced deposition flux near
 descending step edge; In  Fig.5,  the deposition flux averaged over
 the three sites adjacent to the descenting step edge is well matched
 with the aspect ratio for varying deposition angles. As the avearge
 deposition flux near the descending step edge is more enhanced,
 the mean travel length of deposit atoms to ascending step edge
 should become longer, and the growth becomes more unstable giving
 the larger aspect ratio. \par

 For possible origin of growth instability on a vicinal surface,
 two pictures have been proposed based on kinetics of adatoms;
 one attributes the instability to the step Erlich Schwoebel
 barrier effect (SESE) \cite{Bail, Ernst2} and the other to the
 kink Erlich Schwoebel barrier effect (KESE).\cite{vici1, Koponen}
 SESE affects the motion and redistribution of deposited atoms on terrace,
 which should be directly dependent on the deposition flux destribution.
 KESE, however, governs the motion of atoms along step edges,
 and is not directly affected by the initial deposition flux.
 The intimate correlation between deposition flux near descending
 step edge and growth instability shown in Fig.5, indicates that the
 steering-induced deposition flux enhancement near descending step edge
 strengthens the role of SESE on growth instability.\par

%%%%%%Fig. 5%%%%%%%%%%%%%%%%%%%%%%
 \begin{figure}
 \includegraphics[width=0.45\textwidth]{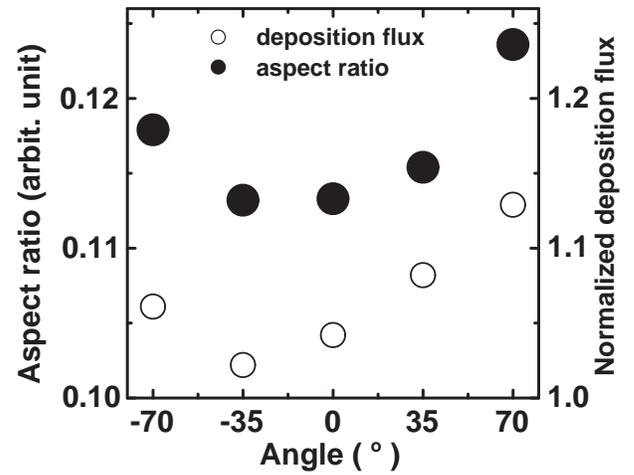}
 \caption{The aspect ratio of lateral roughness to finger width (solid
 circle) and the normalized deposition flux averaged over three
 adsorption sites next to the descending step edge (open circle) are
 plotted as a function of deposition angle. Normalization is made
 with respect to homogeneous flux.}
 \end{figure}
 %%%%%%%%%%%%%%%%%%%%%%%%%%%%%%%%

 In the Figs. 4(a) and (b), the steered growth
 always show larger roughness and smaller coarseness regardless
 of the deposition angle than the growth
 neglecting the steering effect (dotted curves).
 That is, the steering effect always increases
 the growth instability. Such behavior is expected from the relatively
 small flux enhancement near descending step edge for steering-free
 depsoition as shown in Fig.2, consistent with the aforementioned
 explanation.   Although the steering effect is inevitable for vapor
 deposition for thin film growth, the existence of a deposition angle
 producing the minimum growth instability (Fig. 5) suggests
 that the optimizaton of deposition angle should be a prerequisite
 for the most stable growth of thin films on a vicinal surface.\par

 \section{Summary and Conclusion}

 KMC simulation in conjunction with MD simulation is performed to
 study the steering effect, in which the trajectory of each deposit
 atom is affected by interactions with substrate, on  the growth of
 Cu on Cu(1,1,17). It is found that the steered deposition flux
 becomes inhomogeneous and the inhomogeneity depends on the
 deposition angle and direction. The deposition flux enhancement
 near descending step edges is found to be the most critical factor
 for the increase of growth instability due to the steering effect.
 The mechanism of such steering-induced increase of growth
 instability is discussed in details. In the present simulation, we
 also find a deposition angle producing minimum growth instability
 and show that the optimization of deposition angle should be a
desirable for the most stable thin film growth on a vicinal
 surface. \par


\begin{references}

\bibitem{vici0} M. D. Johnson, C. Orme, A. W. Hunt, D. Graff, J. Sunijono,
 L. M. Sander, and B. G. Orr, Phys. Rev. Lett. {\bf 72}, 116 (1994).

\bibitem{Politi} P. Politi, G. Grenet, A. Marty, A. Ponchet, and J. Villain,
 Phys. Rep. {\bf 324}, 271 (2000);

\bibitem{line-by-line} P. Gambardella, A. Dallmeyer, K. Maiti, C. Malagoli,
 W. Eberhardt, K. Kern, and C. Carbone, Nature. {\bf 416}, 301 (2002); P.
 Gambardella, M. Blanc, K. Kuhnke, K. Kern, F. Picaud, C. Ramseyer, C.
 Giradet, C. barreteau, D. Spanjaard, M. C. Desjonqueres, Phys. Rev.
 B{\bf 64}, 045404 (2001).

\bibitem{Bail} G. S. Bales and A. Zangwill, Phys. Rev. B{\bf 41}, 5500 (1990).

\bibitem{Ernst1} L. Schwenger, R. L. Folkerts, and H-J Ernst, Phys. Rev.
 B {\bf 55}, 7406 (1997).

\bibitem{Ernst2} T. Maroutian, L. Douillard, and H.-J. Ernst, Phys. Rev.
 Lett. {\bf 83}, 4353(1999); T. Maroutian, L. Douillard, and H.-J. Ernst,
 Phys. Rev. B{\bf 64}, 165401(2001).

\bibitem{vici1}O. Pierre-Louis and C. Misbah, Y. Saito, J. Krug, and P.
 Politi, Phys. Rev. Lett. {\bf 80}, 4221(1998); O. Pierre-Louis, M. R.
 D'Orsogna, and T. L. Einstein, Phys. Rev. Lett. {\bf 82}, 3661 (1999);
 M. V. Ramana Murty and B. H. Cooper, Phys. Rev. Lett. {\bf 83}, 352(1999).

\bibitem{Koponen} M. Rusanen, I. T. Koponen, J. Heinonen, and T.
 Ala-Nissila, Phys. Rev. Lett. {\bf 86}, 5317 (2001).
 \bibitem{steering1} S. V. Dijken, L. C. Jorritsma, and B. Poelsema, Phys.
 Rev. Lett. {\bf 82},4038 (1999); S.V. Dijken, L.C. Jorritsma, and B.
 Poelsema, Phys.Rev. B {\bf 61}, 14047 (2000).

\bibitem{steering2} J. Seo, S.-M. Kwon, H.-Y. Kim, and J.-S. Kim, Phys.
 Rev. B {\bf 67}, R121402 (2003).

\bibitem{Nissila} J. Merikoski and T.  Ala-Nissila, Phys. Rev. B {\bf 52},
 R8715(1995); J. Merikoski, I. Vattulainen, J. Heinonen, and T. Ala-Nissila,
 Surf. Sci. {\bf 387}, 167(1997).

\bibitem{Sanders} D. E. Sanders and A. E. DePristo, Surf. Sci.{\bf
 254}, 341(1991).

 \bibitem{Zuo} Itay Furman, Ofer Biham, Jiang-Kai Zuo, Anna K.Swan, and
 John F. Wendelken, Phys. Rev.B, {\bf 62}, R10649 (2000).

\bibitem{size-dependence} The flux distribution is determined
 by the shape of potential formed by substrate atoms near surface,
 and  strongly dependent on the terrace width, especially
 for vicinal surfaces with narrow terrace like the present one.

\end{references}
 \end{document}